# Cancer systems biology: exploring cancer-associated genes on cellular networks

Edwin Wang[1]*, Anne Lenferink[2], and Maureen O'Connor-McCourt[2]

1. Computational Chemistry and Biology Group, Biotechnology Research Institute, National Research Council Canada, Montreal, Quebec, H4P 2R2, Canada
2. Receptors, Signaling and Proteomics Group, Biotechnology Research Institute, National Research Council Canada, Montreal, Quebec, H4P 2R2, Canada

*To whom correspondence should be addressed. Email: edwin.wang@cnrc-nrc.gc.ca
Fax: 1-514-496-5143
Tel: 1-514-496-0914

**Abstract**

Genomic alterations lead to cancer complexity and form a major hurdle for a comprehensive understanding of the molecular mechanisms underlying oncogenesis. In this review, we describe the recent advances in studying cancer-associated genes from a systems biological point of view. The integration of known cancer genes onto protein and signaling networks reveals the characteristics of cancer genes within networks. This approach shows that cancer genes often function as network hub proteins which are involved in many cellular processes and form focal nodes in the information exchange between many signaling pathways. Literature mining allows constructing gene-gene networks, in which new cancer genes can be identified. The gene expression profiles of cancer cells are used for reconstructing gene regulatory networks. By doing so, the genes, which are involved in the regulation of cancer progression, can be picked up from these networks after which their functions can be further confirmed in the laboratory.

Cancer is an extremely complex, heterogeneous disease, which could display a degree of complexity at the physiological, tissue and cellular levels. The interactions between tumors and their microenvironments reflect the physiological complexity of cancers, which is the recent focus of the cancer research. Bidirectional interactions between cancer and its microenvironment might promote their growth, survive and the occurrence of distant metastasis [1], However, the molecular mechanisms underlying the interactions between cancer cells and their microenvironment are poorly understood. A cancer tissue or a tumor often contains several distinct pathological featured cancer subtypes, which is recognized as cancer tissue complexity. This tissue complexity is believed to provide functional redundancy for tumors to maintain cellular heterogeneity which could lead to tumor recurrence [2-4] as long as a cancer subtype or a fraction of cancer cells with metastatic potential survives after anticancer treatment. One cancer subtype is able to functionally replace another or even multiple subtypes, which were killed by medical treatments such as using anti-cancer drugs [5, 6]. The functional replacement of cancer subtypes allows tumor survival, further proliferation and finally tumor recurrence. It is reasonable to think that each cancer cell subtype within a tumor might originate through different cancer-specific-developmental mechanisms and mutations in distinct genes. Therefore, this complexity will require a combination of several drugs or treatments targeting various cancer cell subtypes within a tumor. In the past few years, successful progresses have been made to identify the molecular signatures of various cancer subtypes in tumors by performing large-scale gene expression profiling analyses using the microarray technology. For example, sets of gene expression signatures have been identified for breast cancer subtypes [7-9]. Nevertheless, for effective cancer treatment, it is necessary to identify those oncogenic signaling pathways that are the driving force for each of these cancer cell subtypes. However, linking cancer subtypes to the oncogenic signaling pathways and cascades is still hurdled by a poor understanding of the oncogenic processes at the cellular level. The co-existence of several cancer cell subtypes, which rely on the activation of different signaling pathways in one particular tumor, represents the tissue complexity of cancers, while the activation of multiple pathways that lead to the development of the same type of cancer represent the cellular complexity of cancers.

Cancer cells have characteristics of uncontrolled cell growth, the ability to invade their surrounding tissues and finally to generate metastasis in distant places of human body. The accumulation of genetic mutations in part triggers tumor development and progression. Gene mutation or deregulation also promotes cell mobility that is highly correlated with tissue invasion and the formation of distant metastasis. In cancer, many kinds of gene alterations such as gene sequence mutations [10, 11], gene and chromosomal fragment amplifications, chromosomal translocations and gene fusions [12-15], gene deletions [16, 17] and even the mutations and deregulations of noncoding RNAs such as microRNAs [18-20] have been studied and documented extensively. A recent genome-wide screening of cancer mutation genes revealed that different cancer clinical samples of a same cancer type contains different sets of mutated genes which have divergent functions, indicating that the mutated genes do not belong to a same pathway, and therefore, suggesting that a cancer could develop through multiple genetic routes [21]. Because gene activity and regulation that ultimately define a cancer phenotype, it is essential to get a comprehensive understanding of the precise genetic mutations and the consequences of these mutations and genetic alterations. Therefore, it is not surprising that the majority of research efforts are focusing on the genomics, functional genomics and proteomics of cancer cell progression and metastasis.

The complexity of the cancer forms a major obstacle for a comprehensive understanding of underlying molecular mechanisms of oncogenesis. No gene is an island. Even in a single cell, genes are interrelated to work together and take part in many biological processes which then determine the cell's behavior and phenotype. Scientists have struggled for years to figure out how to handle this biological complexity. Systems biology, or more specifically network biology, is driven by the gradual realization that a particular biological function is not from the result of the activity encoded by one single gene. The goal of systems biology is to combine molecular information of various types in models to understand biological systems and their complexity, and finally attempt to predict biological function at the cellular, tissue, organ, and even whole-organism levels. The development of genomic technologies such as high-throughput sequencing, especially DNA, protein microarrays and mass spectrometry, has made it possible to describe cells' biological states in a quantitative manner, and to simultaneously study many gene and protein components and then clarify how these components work together in regulation and carrying out biological processes. The integration of these experimental techniques with the information technology provides a powerful approach to address and dissect the complexity of cancer and other biological problems at various levels in a systems-manner.

**Biological understanding of cellular networks**
In cells, interdependent interactions of genes and proteins form complex cellular networks such as signaling networks, gene regulatory networks and metabolic networks. Cellular networks are the basis of biological complexity. Therefore, the cellular networks have become the core of systems biology. Traditionally, network and graph theory is a branch of mathematics. Here we briefly review and explain network and graph theory with a focus on biological insights. Recent developments in high-throughput techniques in the field of genomics and proteomics research generated vast amount of data, furthermore, electronic format information in literature is becoming accessible on internet. Extraction of these datasets and information used to generate new cellular networks or integrate onto and expand existing cellular networks makes it attractive to study the structures of these networks by relating them to biological properties and insights. Therefore, it is necessary to develop systematic methods for analyzing cellular networks as well as understanding their properties in a cellular context.

In biology, cellular networks include protein interaction networks which encode the information of proteins and their physical interactions, signaling networks which illustrate inter- and intracellular communications and the information process between signaling proteins, gene regulatory networks which describe the regulatory relationships between transcription factors and/or regulatory RNAs and genes, and metabolic networks of biochemical reactions between metabolic substrates and products. Metabolic networks are not the focus in the review, however, more information about metabolic networks can be found in a recent review [22]. Subcellular networks include amino acid residue interaction networks in protein structures, which are assumed to involve a permanent flow of information between amino acids [23].

Networks can be presented as either directed or undirected graphs. Protein interaction networks are modeled as undirected graphs, in which the nodes represent proteins and the links represent the physical interactions between the proteins. Directed graphs, on the other hand, are used to present gene regulatory and metabolic networks. In gene regulatory networks, nodes represent transcription factors or genes, while links represent regulatory relations between transcription factors and the regulated genes or transcription factors. Signaling networks are presented as graphs containing both directed and undirected links. In the networks nodes represent proteins, directed links represent the activation or inactivation relationships between proteins, while undirected links represent physical interactions between proteins. Comparing to other cellular networks, signaling networks are far more complex in terms of the relationships between proteins, for example, nodes may represent different functional proteins such as kinases, growth factors, ligands, receptors, adaptors, scaffolds, transcription factors and so on, which all have different biochemical functions and are involved in many different types of biochemical reactions that characterize a specific signal transduction machinery.

In the past few years, significant progress has been made in the identification and interpretation of the structural properties of cellular networks. This information has shed light on how such properties might reflect the biological meanings and behaviors of cellular networks [24, 25]. Although each type of the cellular network has its own properties, they all share some common structural properties. Cellular networks and other real-world networks, such as a public transportation network exhibit a global structure property that is defined as "scale-free". In a scale-free network, a small group of nodes act as highly connected hubs (high degree), whereas most nodes have only a few links (low degree). For example, a map describing the air transportation in the United States is a network, in which only a few big airports (hubs) in big cities such as Boston, New York, Chicago and Los Angles have many air routes (links) to other airports, while many small airports just have a few air routes to the nearby big airports. This common structural feature encodes a special property of these networks: they are robust but also very vulnerable to failure and attack [25]. In a scale-free network, randomly removal of a substantial fraction of the low degree nodes will make little damage on the network's connectivity, however, targeted removal of the high degree hub nodes will easily disconnect and destroy the network completely, as illustrated by the air transportation map. Disabling big airports (hubs) will wreak havoc in many ways, while damaging a few small airports will have little or no effect on overall air transportation.

In regulatory networks, hub genes are global transcription factors. They may govern a large amount of genes in response to internal and external signals. To fit their multiple biological functions, the hub's expression will have to display dynamic characters. Analysis of the yeast gene regulatory network in which the gene expression profiles of many different cellular conditions were integrated, shows that the hub transcription factors do control a large spectrum of biological processes [26]. We have integrated a genome-wide mRNA decay data onto the *E. coli* gene regulatory network and revealed that the transcription factors whose mRNAs have fast decay rates are significantly enriched in hub genes, suggesting that the expression of the hub genes in gene regulatory networks are indeed highly dynamic. This dynamic behavior facilitates a rapid response of the network to external stimuli [27]. A similar result was obtained in a recent study, in which mRNA decay data were mapped onto the yeast protein interaction network, showed that the hub proteins in protein interaction networks also display fast mRNA decay rates [28]. In protein interaction networks, hub proteins are involved in a large number of interactions, meaning that these proteins will take part in many biological processes and therefore would have higher dynamics in expression. Furthermore, hub proteins may be more important for an organism's survival and have a much broader effect on a system than non-hub proteins. A series of reports confirm this notion [24, 29-32]. These reports also suggest that hub proteins have central positions in cellular networks and are more essential for the organism's survival than other proteins. Therefore, the structure, or in another word, topology, of cellular networks not only sheds light on the complex cellular mechanisms and processes, but also gives insight into evolutionary aspects of the proteins involved. By examining protein evolution and protein interaction networks, Saeed and Deane found that hub proteins are "old" proteins which have evolved more slowly than other proteins [33]. Biologically this makes perfect sense, in that hub proteins are involved in many biological processes and are subject to selection pressure and constraints. Hub proteins in signaling networks are the focal nodes that are shared by many signaling pathways. In another word, hub proteins have become information exchanging and processing centers. Alterations to these hub proteins may therefore globally affect the well being of living cells. A recent RNAi screening of worms supports this hypothesis. Lehner et al. systematically mapped the genetic interactions of *Caenorhabditis elegans* genes involved in signaling pathways and revealed a network of 350 interactions [34]. They then tested 65,000 pairwise gene interactions and found that a few genes interact with an unexpectedly large number of signalling pathways. These hub genes were identified as chromatin-modifying proteins which are conserved across animals where they display core genetic buffering properties.

Cellular networks are complex systems, in which a gene does not independently performing a single task, instead, individual genes can be grouped, which collaborate to carry out some specific biological function. We call such a gene group as a functional module. This assumption leads to the idea that a complex network can be broken up into many small but functional modules or units, which can be then studied to determine their structural properties and functional behaviors. Once we understand the functions, properties and regulatory/interaction behaviors of these modules, we can then use these functional modules to rebuild sub-networks and even whole networks and study their properties and functions. Network motifs are examples of such functional modules, which are the statistically significant recurring structural patterns or small subgraphs or sub-networks that are found more often in a real network than would be expected by chance [35]. These motifs are known as gene regulatory loops in biology. These motifs can self-organize or are forming a network by sharing nodes between various motifs [27]. Network motifs have been studied in details in gene regulatory networks. Three major motifs are found in gene regulatory networks: Single Input Module (SIM), bi-fan and Feedforward Loop (FFL) (Figure 1). One design principle of these motifs is that the transcription factors whose mRNAs have fast decay rates are significantly enriched in these motifs, suggesting that motif structures encode a regulatory behavior: network motifs are able to rapidly response to internal and external stimuli and decrease cell internal noise [27]. Network motifs have been shown to have distinct regulatory functions and are robust in that they are resistant to internal noise. Both theoretical and experimental studies have shown that network motifs bear distinct regulatory functions and particular kinetic properties that determine the temporal program of gene expression [36]. Therefore, the frequencies and types of network motifs with which cells use reveal the regulatory strategies that are selected in different cellular conditions [27, 37, 38]. For example, FFLs are buffers that respond only to persistent input signals [39], which makes them well-suited for responding to endogenous conditions, while the motifs whose key regulator's transcripts have a fast mRNA decay rate are preferentially used for responding to extraneous conditions [27]. In signaling networks, network motifs such as switches [40], gates [41], and positive or negative feedback loops provide specific regulatory capacities in decoding signal strength, processing information and controlling noise [42, 43].

Distinct network motifs could form large aggregated structures, called network themes that perform specific functions by forming collaborations between a large number of motifs [44]. In this case, network themes can be regarded as communities of functionally related nodes. A large protein complex in protein interaction networks is one of the examples of such network community.

**Integrative network analysis of cancer-associated genes**
High-throughput gene expression profiling often leads to the identification of a hundred or sometimes even thousands of modulated genes for a given phenotype. However, the extraction and interpretation of biological insights of the differientially expressed genes in these high-throughput datasets are challenging, and limited by the difficulties in recognizing the gene-gene relations and associations within the huge amount of data. Although it is possible to classify the identified genes into different functional groups using Gene Ontology (GO) [45], the in-depth relationships between genes in different functional categories can still not be easily illustrated. A particular phenotype is the result of collaborations of a group of genes, which are not necessary belonging to one same functional category. Therefore, integration of microarray generated gene lists onto cellular networks could help analyzing and interpreting the biological significance of the genes in a network and their gene-interdependent context. This notion provides a structured network knowledge-base approach to analyze genome-wide gene expression profiles in the context of known functional interrelationships among genes, proteins and phenotypes.

Motivated by this concept, Wachi et al. investigated the differentially expressed genes in squamous cell lung cancer which were identified by projecting the microarray gene expression profiling onto a human protein interaction network [46]. The data for the network construction were taken from the online predicted human interaction database, (OPHID) [47], which contains 16,034 known human protein interactions obtained from various public protein interaction databases, and 23,889 additional protein interactions that were predicted. They mapped the 360 up-regulated and 270 down-regulated genes that were identified in the lung cancer microarray experiment onto the protein interaction network. Further network analysis revealed that the up-regulated genes in this dataset are well connected, whereas the suppressed genes and randomly selected genes are less so. They also showed that high degree of centrality in these differentially up-regulated genes, but not for the genes that are suppressed. These results imply that the up-regulated, but not down-regulated genes in this experiment are enriched in hub proteins, which are associated with essential functions in protein interaction networks [29]. Cancer cells are characterized by uncontrolled growth, which could suggest that the induced genes in cancer cells, compare to normal cells, are more essential for survival and proliferation. The work described here uncovers the characteristics of cancer-associated genes in a network context and supports the notion that integrative network analysis of large datasets obtained from gene expression profiling helps understanding the functions of biological systems.

    The characteristics of cancer-associated genes uncovered in this study were confirmed by a recent analysis of a human protein interaction network integrated with literature-mined cancer genes. Johsson and Bates [48] used mutated cancer genes collected from literature [49] and attempted to uncover their intrinsic properties in a human protein interaction network which was constructed from the entire human genome using an orthology-based method [50]. In total, 346 genes encoding 509 protein isoforms, were mapped on to the network. This analysis showed that cancer proteins have on average, twice as many interaction partners as other proteins in the network, which implies the evolutionary aspects of cancer genes. Accumulating evidence shows a positive correlation between the evolution of proteins and their number of interactions within a given network [31, 51, 52]. With this consideration in mind, the authors concluded that proteins, whose mutation results in a detrimental change of function that leads to cancer, may generally be more conserved than other proteins. Alternatively, cancer proteins, as they have more interaction partners, may be involved in significantly more biological processes and play a central role in the protein network. To further explore this direction, Johsson and Bates also investigated the relationships between these cancer genes and network communities, which represent a distinct biological process, meaning that if a protein is a member of multiple network communities, it takes part in more biological processes. The results of this analysis show that the identified cancer proteins are indeed involved in more network communities than other proteins in the network, suggesting their more prominent centrality and participation in the formation of the proteome network backbone. Taking it one step further, the authors also analyzed the domain compositions of these cancer proteins. Cancer proteins display a high ratio of highly promiscuous domains, in terms of the number of different proteins with which they interact, indicating that they play central roles in many biological processes and that mutations in these proteins could lead to a higher cancer incidence. Moreover, the domains most frequently found in the cancer protein population have functionalities that particularly focus on DNA regulation and repairing, such as Zinc-finger, PHD-finger, BRCT and Paired-box domains, which all happen to be transcription factors.

    These works provide a biological insight into the global protein interaction network properties of cancer proteins and uncover one of the most striking properties of cancer proteins in that cancer-associated proteins are network hubs, which play central roles in biological systems and take part in many biological processes. Taken together, each hub cancer protein may reflect a specific domain of a cellular function, which suggests that mutations of an individual or a few hub proteins together may lead to oncogenesis or cancer progression. However, these studies provide little insights into the oncogenic mechanisms simply because protein interaction networks have limited information compared to signaling networks in which protein regulatory (activation and blocking) information is encoded. Therefore, the integration of cancer genes onto the

existing and established signaling networks would be possible to get more insights into the oncogenic process and cancer progression.

Cells use a sophisticated communication between proteins to perform a series of tasks such as growth and maintenance, cell survival, apoptosis and development. Signaling pathways are crucial to maintain cellular homeostasis and determine cell behavior. Therefore, alterations in the expression of genes and their regulators will reflect on these cellular signaling pathways which in turn lead to tumor development and/or the promotion of cell migration and metastasis. In deed, mutations in genes that encode signaling proteins are commonly observed in many types of cancers [53].

Specific signaling pathways deploy many different proteins, however, pathways often "talk" each other. This so called "cross-talk" between pathways has been systematically investigated in a recent study, and an unexpected high numbers of cross-talk events between signaling pathways were discovered [54]. These results indicate that signaling pathways form a complex network to process information. Structural analysis of a literature-mined human cellular signaling network containing ~500 proteins, showed that signaling pathways are intertwined in order to manage the numerous cell behavior outputs [55]. This work provides a framework for our understanding of how signaling information is processed in cells. Furthermore, analysis of interactions between microRNAs and the same signaling network reveals the principles of microRNA regulation of the network [56]. Together, these approaches hint that an integrative analysis of signaling networks with cancer proteins would highlight the characteristics of cancer proteins within these signaling networks.

Errors in signal transduction can lead to altered development and incorrect behavioral decisions which could result in uncontrolled cell growth or even cancer. The relationships of signaling proteins are thought to be critical in determining cell behavior, therefore the mapping of cancer genes on the nodes of a signaling network could general lead us to which mechanisms support the continued survival and proliferation of cancer cells. We manually curated human cellular signaling pathways and merged these curated data into another literature-mined human cellular signaling network mentioned previously [55]. As a result, the new network contains ~1,100 signaling proteins. Next the cancer proteins were obtained from NCBI's Online Mendelian Inheritance in Man (OMIM) database [57] were mapped onto the network. Nearly 90 cancer proteins were mapped onto the network [58]. Not surprisingly, cancer proteins are enriched in hub proteins in the signaling network. As mentioned, cancer genes often get mutated, which could result in the activation of particular focal signaling nodes that play important roles in the information exchange between many individual signaling pathways. Indeed, several cancer proteins form the focal nodes in signaling networks and therefore play important roles in cancer development.

The cellular signal information flow initiates from the extracellular space, e.g. a ligand binds to a cellular membrane receptor to generate the signal that is then transmitted by intracellular signaling components in cytosol to the signaling components within nucleus. This process of signal transduction is sensitive in terms of mutated genes, which result in altered signaling and therefore tumourgenesis, and increase cell mobility and invasion. We found that cancer proteins are enriched in the downstream section of signaling networks, the realm of the transcription factors [58]. Along with this discovery, we also found that cancer proteins are hardly represented in some particular network motifs such as bi-fan (Figure 1), which is a structure with regulatory redundancy but also one of the most abundance network motifs in central region of the human signaling network. These results lead us to believe that the central region of a signaling network provides a genetic buffer for cells in that it may prevent cancer development, which is in agreement with the robustness of networks [59]. The fact that cancer proteins are enriched in the downstream region hints that proteins in this region are crucial for determining specific cell behavior. Our work provides insights into the signaling networks invoked in cancer development and progression.

The systems-level approach taken in these works, i.e. combining information on how proteins interact with each other and how transmitted signals are processed, with information on known cancer genes and gene expression in cancer cells, is a particularly appealing approach to gain an understanding of complex biological processes, such as cancer development and metastasis. Network analyses using comprehensive knowledge of biology provide a framework for structuring the existing knowledge regarding cancer biology and help identifying proteins and/or significant functional modules and the underlying mechanisms of the oncogenic process.

**Hunting new cancer-related genes using cellular networks**
Protein interaction networks have been used to hunt new cancer-associated genes. Jonsson et al. have been motivated to find genes involved in metastasis by integrating cancer cell microarray expression data onto a rat protein interaction network which was constructed by transferring protein-protein interaction information from other species using the protein homology concept [50]. The network was evaluated by a confidence scores based on their homology to proteins that have been experimentally observed to interact. Metastasis is a key event that is usually associated with a poor prognosis in cancer patients. Metastasizing cancer cells have special properties, in that they can display features such as increased motility and invasiveness.

It was hypothesized that sub-networks of protein interactions may govern the metastasis. Jonsson et al. used a data set containing the up- and down-regulated genes that was obtained from a cancer microarray study, and constructed sub-networks around proteins which was then evaluated using cluster analysis to define network communities that reflect small protein interaction units that are involved in the metastastic process [50]. As a result, they identified 37 protein communities of highly interconnected proteins, and of which most of them have been associated with cancer and metastasis.

Gene networks have been constructed by merging various data sources which were then used to find or prioritize cancer and other disease genes. In this context, gene-gene networks are presented using undirected graphs, in which the nodes represent genes and the links represent relations between genes. The relations of the genes can be protein physical interactions, gene regulatory relations, and gene associations and so on. Franke et al. constructed such a human gene-gene network using the databases of known interactions, GO, microarray co-expressions and yeast two-hybrid data [60]. They then integrated this network with already known genetic information of diseases (i.e. genetic loci for a particular human disease). The authors reasoned that the cancer genes from each locus are likely to be involved in one same molecular pathway and biological process. To prove their concept, they showed that the genes prominent in any one disease were closer to each other in the network than would be expected by chance, which suggests that these genes are involved in the disease and therefore tend to have more functional interactions or associations. To assess the predictive power of this method, the authors tested it by picking disease genes using the network. Four out of 10 breast cancer genes were ranked in the top of the gene list, which is 4 times higher than a breast cancer gene that would be picked by chance. When they integrated more interaction data onto the network and adjusted the network topology, the ranking of these disease genes improved considerably, and included 9 of the 10 genes. These results indicate that the use of a network significantly improves the chance of finding the correct cancer genes.

In the past few years, a series of studies focused on constructing gene-gene networks using data from literature and other sources. One notion behind this is that nearly 80% of biological information and data are coded in natural language in technical reports, web sites, research publications and other text documents [61]. To facilitate the extraction of these data, methods have been developed for the automatic extraction of interaction and pathway information from the scientific literature [62-66]. Furthermore, the extracted relations between genes have been used to construct gene-gene networks, and several software packages and related datasets have been developed. PubGene [67] is an example of such a tool, which contains a database and analysis software for constructing gene-gene networks by identifying relationships between genes based on their statistical co-occurrence in the abstracts of scientific papers. The Information Hyperlinked over Proteins (iHop) [68] is another example. In this case one can use gene names to retrieve gene-gene relations from PubMed abstracts that match a specified gene/protein name. iHop also provides automatic extraction gene-gene relations for software developers and bioinformatics scientists.

Contrast to most of the text mining methods that use the abstracts of research papers, Natarajan et al. tried to use full-length scientific articles to extract gene-gene relations [69], and also fused the extracted gene interactions to structured data and knowledge bases such as Ingenuity Pathway Analysis, UniProt [70], InterPro [71], NCBI Entrez and GO. A human gene-gene network was constructed using theses data sources. The authors then mapped the differentially expressed genes identified from microarray, which profiled the gene expression in glioblastoma as a response to S1P in vitro. Further analysis led to the identification of a cascading event that is triggered by S1P, and which leads to the transactivation of MMP-9 via neuregulin-1, vascular endothelial growth factor, and the urokinase-type plasminogen activator. This suggests that the interaction network has the potential to shed new light on our understanding of the cancer-related process. Therefore, automated extraction of information from biological literature, together with combining and integrating biological data from laboratory experiments, provides an effective way in biological knowledge discovery.

**Reverse engineering of gene regulatory networks from microarray data**
Reverse engineering of biological networks is a process of elucidating the structure of gene regulation relationships by reasoning backwards from the observations of gene expression values. In recent years, a substantial effort has been made to reconstruct gene regulatory networks using microarray profiles. Here we just describe two related work which combined computational and experimental approaches.

Basso et al. developed a statistical algorithm using mutual information for more accurately reasoning networks in which pair-wise gene-gene interactions are described [72]. The algorithm was named the Algorithm for the Reconstruction of Accurate Cellular Networks (ARACNE). To test ARACE, the authors used a huge number of gene expression profiles (336 samples) of human B-cell at different stages covering normal to cancer cells to construct a network. A sub-network was used for validation using GO and chip-on-chip experiments. The results are encouraging in that 90% specificity was obtained for ARACNE. However, we should be aware that the test did not include the predictions with lowest mutual information scores. Nevertheless, this approach shows that with "enough" gene expression data, reasonable gene networks can be retrieved by developing proper algorithms.

Another example of the reverse engineering applied to cancer research was carried out using a dataset that was generated in our own laboratory. We constructed a gene regulatory network using the time course microarray profiles from a mouse epithelial breast cell line (BRI-JM01), which was isolated from mammary tumors in transgenic mice. These cells undergo an epithelial to mesenchymal transition (EMT) when they are treated with TGF-β [73]. To identify the transcriptional changes underlying this EMT, we exposed the BRI-JM01 cell line to TGF-β for 7 time intervals (0.5-24 h), and interrogated the transcriptome using cDNA microarrays. Based on the microarray profiles and the markov chain based network construction method [74], we constructed a gene regulatory network that contains nearly 50 genes and 3 layers of regulations, in which the regulatory relations are either direct or indirect (Lenferink et al., unpublished data, Figure 2). Known biological information was used to validate the network. Interestingly, in the top layer of the network, all the annotated genes are either transcription factors or signaling proteins, which are known as regulatory proteins. Most known genes in the bottom layer of the network are known to involve in cancer processes, which suggest that the network somewhat seems right. Notably, clusterin, one of the genes that are up-regulated in the middle and late time-points shows many regulatory links to other genes in the network. During the EMT process, clustrerin is secreted by the BRI-JM01 cells. Interestingly, when applying anti-clusterin antibodies to the TGF-β treated BRI-JM01 cells, we were able to block the TGF-β induced EMT. This result strongly implies that the secreted form of clusterin plays a pivotal role in the TGF-β induced EMT and therefore TGF-β's tumor promoting effects on the BRI-JM01 cell line. Currently, reverse engineering of gene regulatory networks using microarray data is mainly hurdled by limited microarray experiments we could perform for a given sample. Reverse engineering methods only provide some hints to biologists, although they could narrow down the gene list of interest. A substantial lab experiments should be followed to further validate the genes of interest from the inferred gene regulatory networks.

**Outlook**
The analysis of the cancer phenomenon using a systems-level approach is still in its infant shoes. New and emerging technologies need to be developed and validated.
These technologies include single cell signaling mapping, which will be very helpful in obtaining the full picture of signaling dynamics occurring in different cancer cells and during various stages of cancer development. These techniques will be especially useful for understanding the biology of tumors which consist of notoriously heterogeneous cancer cell populations. The information of relations between genes and/or proteins is still limited, but will be alleviated once new high-throughput datasets become available.

These new datasets either generated experimentally or by literature mining will no doubt provide information on new interactions between genes. Current efforts are ongoing to curate high quality signaling data from literature [75, 76].

Overall, the systems biology output will bring unprecedented amounts of molecular information and large-scale datasets to medicine in the form of DNA sequences and quntative information of mRNAs, proteins, and metabolites. An important part of systems biology is taking all of these measurements in consideration to construct models to describe what is going on in a cell, a tissue, an organ, or even an organism. A systmes-level understanding the underlying mechanism causing cancer in an individual cancer patient will allow science to become more focused and will contribute significantly to the clinical application of a personalized medication.

Figure legends:

Figure 1. Network motifs in gene regulatory networks.
Nodes represent genes and lines represent gene regulatory relations. **A,** Single Input Module (SIM): a transcription factor (TF) regulates a group of genes (G1, G2, G3 and G4). **B,** Feedforward Loop (FFL): a transcription factor (TF1) regulates the second transcription factor (TF2), both TF1 and TF2 regulate a target gene (G1). **C,** Bi-fan: both transcription factors TF1 and TF2 regulate both target genes (G1 and G2).

Figure 2. A gene regulatory network inferred from the time course gene expression profiles of BRI-JM01 cell line. Nodes represent genes and lines represent gene regulatory relations.

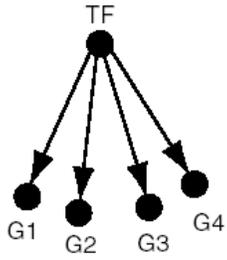 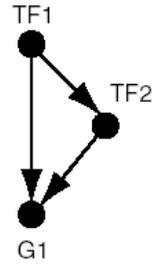 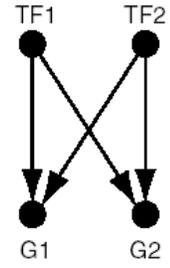

Figure 1

Figure 2